\documentclass[prl,preprint,showpacs]{revtex4}

\usepackage{graphicx}

\begin{document}

\title{$c$-axis transport and phenomenology of the pseudo-gap state  
in $Bi_2Sr_2CaCu_2O_{8+\delta}$}

 \author{M.Giura}
 \affiliation{Dipartimento di Fisica and Unit\`{a} INFM,\\
 Universit\`{a}
 ``La Sapienza'', P.le Aldo Moro 2, 00185 Roma, Italy}
 \author{R.Fastampa}
 \affiliation{Dipartimento di Fisica and Unit\`{a} INFM,\\
 Universit\`{a}
 ``La Sapienza'', P.le Aldo Moro 2, 00185 Roma, Italy}
 \author{S.Sarti}
 \affiliation{Dipartimento di Fisica and Unit\`{a} INFM,\\
 Universit\`{a}
 ``La Sapienza'', P.le Aldo Moro 2, 00185 Roma, Italy}

 \author{E.Silva}
 \affiliation{Dipartimento di Fisica ``E.Amaldi'' and Unit\`{a}
 INFM,\\ Universit\`{a} ``Roma Tre'', Via della Vasca Navale 84, 00146
 Roma,
 Italy}

\date{\today}

\begin{abstract}

We measure and analyze the resistivity of $Bi_2Sr_2CaCu_2O_{8+\delta}$ crystals for different doping $\delta$. We obtain the fraction of carrier $\eta(T,\delta) = n_g/n_{TOT}$  that do not participate to the c-axis conductivity. All the curves $\eta(T,\delta)$ collapse onto a universal curve when plotted against a reduced temperature $x=[T-\Theta(\delta)]/\Delta^{*}(\delta)$. We find that  at the superconducting transition $n_g$ is doping independent. We also show that a magnetic field up to 14 T  does not affect the degree of localization in the (a,b) planes but widens the temperature range of the x-scaling by suppressing the superconducting phase coherence.

\end{abstract}

\maketitle

\begin{figure}
\includegraphics[scale=.55,angle=0]{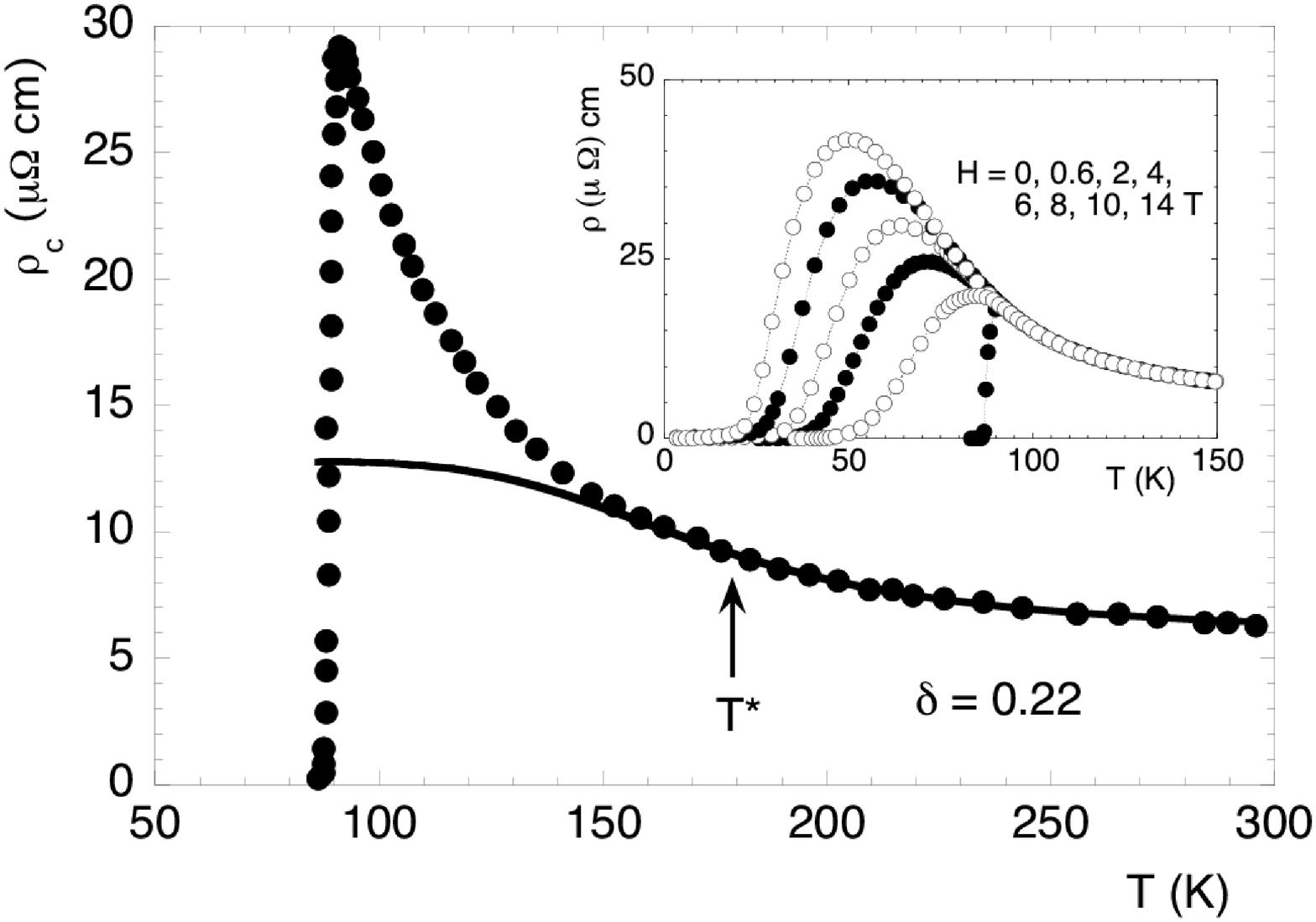}
\caption{Temperature dependent c-axis resistivity for the sample with $\delta = 0.22$ and $H=0$. The solid line is the fit according to the model reported in Ref. \cite{giura}. Inset: resistivity data for the sample with $\delta = 0.235$ for different applied magnetic fields.}
\label{fig1}
\end{figure}

\begin{figure}
\includegraphics[scale=1,angle=0]{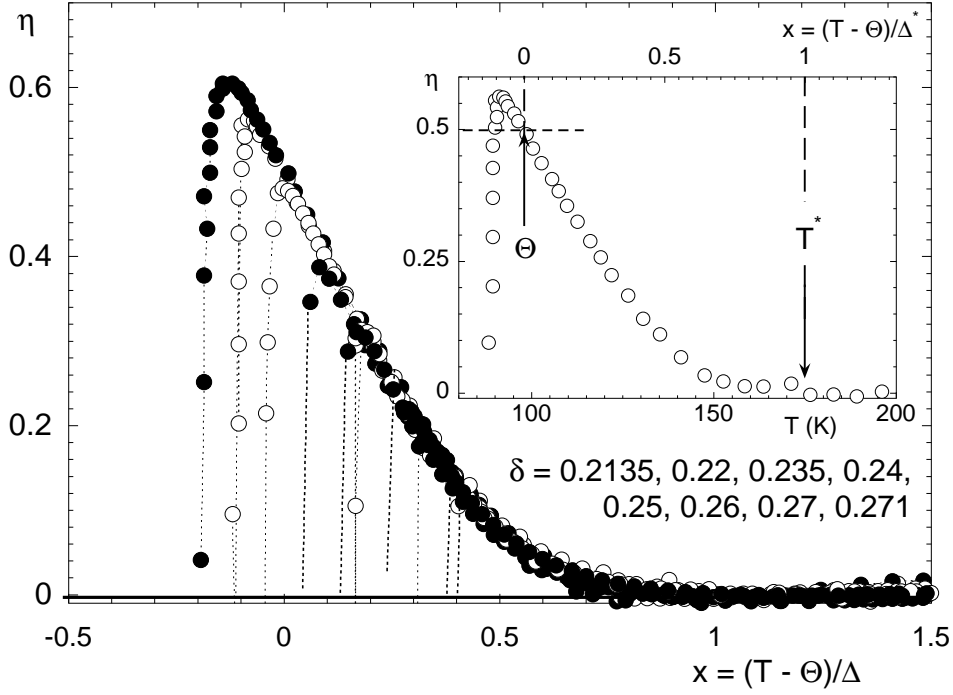}
\caption{Universal behavior of $\eta$ vs $x$ (see text). In the insert, the curve $\eta(T)$ and the operative definition of $\Theta$ and $T^*$ for the sample with $\delta=0.22$ (see also Fig.\ref{fig1}).}
\label{fig2}
\end{figure}

\begin{figure}
\includegraphics[scale=1,angle=0]{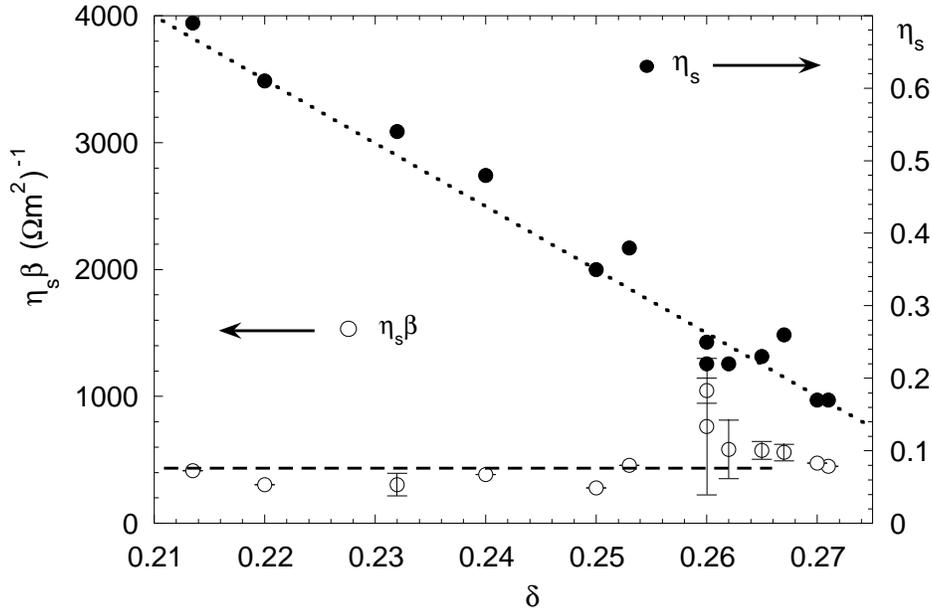}
\caption{Doping dependence of $\eta_{s}$, the gapped fraction at $T_c$ and of the the product $\eta_{s}\beta$. The number of gapped carriers at $T_c$, $n_g \propto \eta_s\beta$ is found to be constant as a function of $\delta$, up to $\delta = 0.27$. Error bars are given by the uncertainity in the value of $\beta$ (see Ref. \cite{giura})}
\label{fig3}
\end{figure}

A comprehensive description and a detailed explanation of the
experimental behavior of the conductivity tensor is an unavoidable
step for the understanding of the physics of cuprate superconductors.
The doping as well as the temperature dependences of the
in plane ($\rho_{ab}$) and out of plane ($\rho_c$) resistivities above
the superconducting critical temperature remain a vigorously debated
topic.  It is commonly accepted that those dependences contain most,
if not all, of the peculiarities of the cuprate superconductors.  In
particular, the temperature dependence of $\rho_{ab}$, with the
apparent absence of features in a very extended temperature range, was
taken as a first signature of a non-Fermi-liquid
state \cite{nonFermi}.\\
It is believed that a full comprehension of the normal state transport cannot be obtained without considering both $\rho_{ab}$ and $\rho_c$. \cite{nonFermi}
Many papers \cite{kumar,altri} were devoted to look for some 
relation between these two quantities.  In 
Bi$_2$Sr$_2$CaCu$_2$O$_{8+\delta}$ (Bi2212), owing to its unambiguously
layered structure, the out of plane $\rho_c$ resistivity has been 
described by means of tunneling phenomena along the $c$-axis, induced 
by the two-dimensional character of the CuO$_2$ planes.  In addition, 
it has been theoretically proposed that, by lowering the temperature 
below a not sharply identified temperature $T^*>T_{c}$, strong 
electronic correlation would set in, leading to complex phenomena 
such as striped phase \cite{stripes}, charge spin 
separation \cite{chargesep}, and others, to all of which we will refer 
in the following as the pseudogap state, whose influence on transport 
properties is not fully and unambiguously understood at present. Also under present discussion is the role of the superconducting fluctuations near $T_c$. \cite{flucts}  In 
any case, independently of their origin all these phenomena are 
expected to further increase the charge confinement within the 
CuO$_2$ conducting plane, as empirically deduced by the increase of 
$\rho_c$ and decrease of $\rho_{ab}$ below $T^{*}$.\\
A description of the conductivity in the pseudogap state requires the knowledge of the conductivity above $T^{*}$ (in the following, we will refer to this region as the ``normal state'').  In a previous paper \cite{giura} we have presented a systematic  investigation of the temperature and doping dependence of the  $c$-axis resistivity in Bi2212.  In the normal state region ($T>T^*$) those data, together with the systematic data of Watanabe et al, \cite{watanabe} were successfully interpreted by means of a simple phenomenological model based on the existence of two energy barriers per unit cell, along the $c$ axis.
The hypothesys that the two layers within the CuO$_2$ bilayers could be considered as two separate layers has been found relevant for the explanation of experimental results in various compounds, \cite{pomar} and for important theoretical predictions. \cite{Leggett}
According to our model, for each barrier the electrical transport  along the $c$ axis is determined by two distinct processes: {\it (i)}  the thermal activation across the barrier and {\it (ii)} the  tunneling through the barrier.  The latter mechanism was identified  with an incoherent process due to the in-plane phonon scattering, as  in the Kumar model \cite{kumar}. This fact implies some relation between the in-plane and the out-of-plane resistivities. The overall expression for the normal state resistivity along the c-axis
comes out from the series of the two barriers:
\begin{equation}
\rho_{c,n} = \frac{1/d_1}{\frac{t_{c,1}^2}{a \rho_{ab,n}+b}+\beta
e^{-\Delta_1/K_BT}}+\frac{1/d_2}{\frac{t_{c,2}^2}{a
\rho_{ab,n}+b}+\beta e^{-\Delta_2/K_BT}}
\label{oldrho}
\end{equation}
where $\rho_{ab,n}$ is the measured normal state in plane resistivity, $a = \frac{\hbar^2n_0}{2gm^*}$ ($n_0$ is the density of charge carriers, $g$ is the density of states for unit area in the \{a,b\} plane), $b$ is a coefficient proportional to the difference between the effect of disorder on the out-of-planes and in-plane resistivity, the prefactor $\beta$ in the activated terms is proportional to $n_0$ and, for each barrier ($i$ = 1, 2), $d_i$ is the spacing between the layers, $t_{c,i}$ is the tunneling matrix element, and $\Delta_i$ is the height of the barrier. Further details are extensively reported in Ref. \onlinecite{giura}.\\
This model was able to successfully describe the experimental $\rho_{c}(T)$ data for large variations of the doping $\delta$, in the temperature range $T>T^*$.  Excellent fits have been obtained in the underdoped region as well as in the slightly overdoped one.  The doping dependence of the resulting parameters were found to be consistent with the general implications of the model. In particular, $g$ was found to be independent on doping, as expected for a 2D system, while $\beta \propto n_{0}$ increases with increasing $\delta$. All these results have been found to be valid up to $\delta \simeq 0.27$. Above this threshold, $g$ starts to increase suggesting a crossover toward a $3D$ state. This is also confirmed by the fact that the parameter $b$ tends to vanish at $\delta \simeq 0.31$, indicating that at high doping the effect of disorder is the same in all directions.\\
In this paper, assuming on the basis of the previous findings that the two barriers model is a correct description of the normal state, we analyze the data of the resistivity in Bi2212 crystals for $T<T^*$, spanning a doping range from underdoped to overdoped.  The results of the analysis give, on one side, a further confirmation of the accuracy of the model and on the other, novel indications on the phenomenology in the temperature region between $T^*$ and the superconducting transition temperature $T_{c}$.  In addition, in order to gain more detailed insight on the pseudogap state, we will 
extend the measurements of $\rho_c$ to the case of an applied magnetic field up to 14 T.  The measurements demonstrate that the magnetic field destroys the superconducting coherence, leaving almost unaffected the pseudogapped state.\\
We focus the attention on the temperature region corresponding to the pseudogap state, $T<T^*$, characterized by the following experimental evidences: (1) The presence of a pseudo-gap ($\Delta_{PG}$) in the electronic spectrum, whose amplitude is temperature independent, as obtained from photoemission spectroscopy \cite{photoemission} and optical conductivity \cite{optical}. (2) The $c$-axis optical conductivity, in the pseudogap region, decreases continuously by lowering the temperature \cite{optical}.\\
In order to find quantitative information from the resistivity data, we adopt the following framework: below $T^*$, some charge carriers form "gapped" states with energy gap $\Delta_{PG}$ (independent on $T$), and do not participate in the $c$-axis conductivity. The fraction $\eta(T)$ of such carriers increases continuously with lowering $T$, progressively reducing the overall number of carriers within the pseudogap.  Being interested in the phenomenology  of the gapped state, we will not address the issue of the microscopic nature of these gapped states (preformed pairs, localized states etc.). Within this frame, the overall $c$-axis conductivity, $\sigma_{c}$, is given by:
\begin{eqnarray}
\nonumber
\sigma_{c} = \sigma_{c,n} \left[1-\eta(T)\right]\\
\label{rhocvseta}
\rho_{c} = \frac{\rho_{c,n}}{1-\eta(T)}
\end{eqnarray}
where $\eta(T \geq T^*)=$ 0 by definition and, in general, $\eta(T)\leq$ 1. Within the present model, the measured resistivity $\rho_{c,exp}$ is given by $\rho_{c}$ for $T>T_c$,  and it is clear that the main physics of the pseudogap region is contained in the parameter $\eta$ and in its temperature and doping dependences. At each doping level $\delta$ the determination of $\eta(T)$ from the experimental data follows a straightforward procedure.  Having already obtained the normal state resistivity, $\rho_{c,n}$, from the fits of the data presented in Ref.\cite{giura}, from Eq.\ref{rhocvseta} one immediately gets \cite{notalinear} $\eta=1-\rho_{c,n}/\rho_{c,exp}$.
The results of the procedure are presented in Fig.\ref{fig1} for an underdoped sample ($\delta=$0.22), where we report the experimental data and the curve fitting the normal state. In the same figure (inset) we report data as a function of the applied magnetic field for the sample with $\delta = 0.235$. The 
``gapped fraction'' $\eta (T)$ is reported in the inset of Fig.\ref{fig2}.  A clear interest resides in the 
doping and temperature dependence of the gapped carrier fraction.  We have then obtained the curves $\eta(T)$ for all our samples and for the measurements reported by Watanabe at al, \cite{watanabe} thus  exploring a rather large doping interval: 0.21 $<\delta <$ 0.27.\\
We now show that all the $\eta(T;\delta)$ collapse on a single curve, when plotted against a properly chosen reduced temperature.  To this aim, we choose as a reference the curve $\eta(T)$ for the sample with $\delta=$0.22. In order to perform a scaling, we need to define, for this reference curve only, the characteristic temperature $\Theta$ as $\eta(\Theta)=\frac{1}{2}$ and the characteristic temperature scale $\Delta^*=T^*-\Theta$. We then define the scaling variable $x=\frac{T-\Theta}{\Delta^*}$.  Once the reference curve $\eta(x)$ is obtained for $\delta=0.22$, the scaling of all the other curves is accomplished by looking for two doping dependent scaling parameters $\Theta(\delta)$ and $\Delta^*(\delta)$ such that for the given doping $\delta$ the curve $\eta\left(\frac{T-\Theta(\delta)}{\Delta^*(\delta)}\right)$ collapses onto the reference curve.
It is an important result of this paper that, for each doping level, such a pair of parameters $\Theta(\delta)$, $\Delta^{*}(\delta)$ exists. The result of this scaling procedure is plotted in Fig.\ref{fig2}.  It clearly indicates that, within the general framework here presented, the fraction of the carriers that do not contribute to the conduction along the $c$-axis (and thus are strongly localized in the CuO$_2$ planes, where the conductivity in its turn increases) arrange themselves on a universal curve as a function of the reduced temperature $x$.\\
Once the "gapped carrier" fraction is obtained as a function of $T$, it is interesting to analyze how this behavior reflects on the superconducting properties.  We define $\eta_{s}=\eta(T_c)$, the value of the gapped fraction at the superconducting transition temperature. It comes out that  $\eta_{s}$ is doping dependent: with decreasing doping a larger fraction of gapped carriers builds up before the superconducting transition occurs.  This is seen in Fig.\ref{fig3}, where we report $\eta_s$ as a function of $\delta$.  In the same figure we report $\eta_{s}\beta$ vs. $\delta$.  It comes out that, within the scattering of the data points, this product is constant up to $\delta = 0.27$.  Since $\beta d exp(-\Delta/K_BT) = \sigma^{th}$, the thermal hopping conductivity, and $\sigma^{th}$ is proportional to the number of charge carriers, \cite{giura} this finding has the important meaning that the number of gapped electrons at the critical temperature $T_c(\delta)$, $n_g\propto \eta_s\beta$, is doping independent up to the overdoped region. This fact suggests that the number of 'gapped' electrons are strictly related to the onset of superconductivity. Following this suggestion, we interpret the 'gapped' electrons as preformed pairs, that condensate into the superconducting state as their number reaches a well defined value $n_g$. Within this interpretative frame, as the temperature is lowered some pairs develop over a temperature range whose scale is given by $\Delta^*$. $\Delta^*$ is a measure of the crossover region between the normal state and the fully paired state.

\begin{figure}
\includegraphics[scale=.6,angle=0]{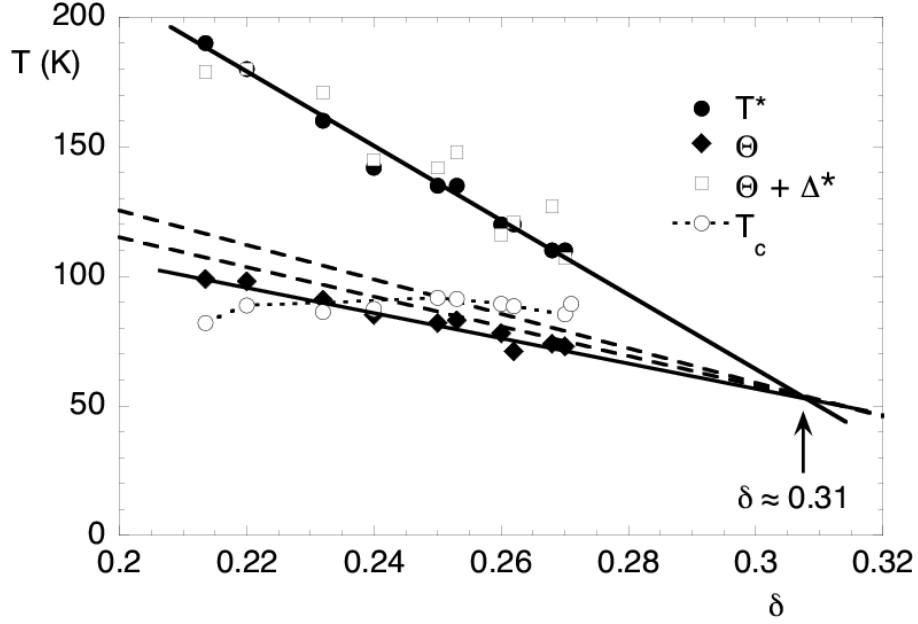}
\caption{Doping dependence of the characteristic temperatures $T^{*}$, $\Theta$, $T_{c}$. Dashed lines correspond to different thresholds for the definition of $\Theta$ (see text).}
\label{fig4}
\end{figure}

\begin{figure}
\includegraphics[scale=1,angle=0]{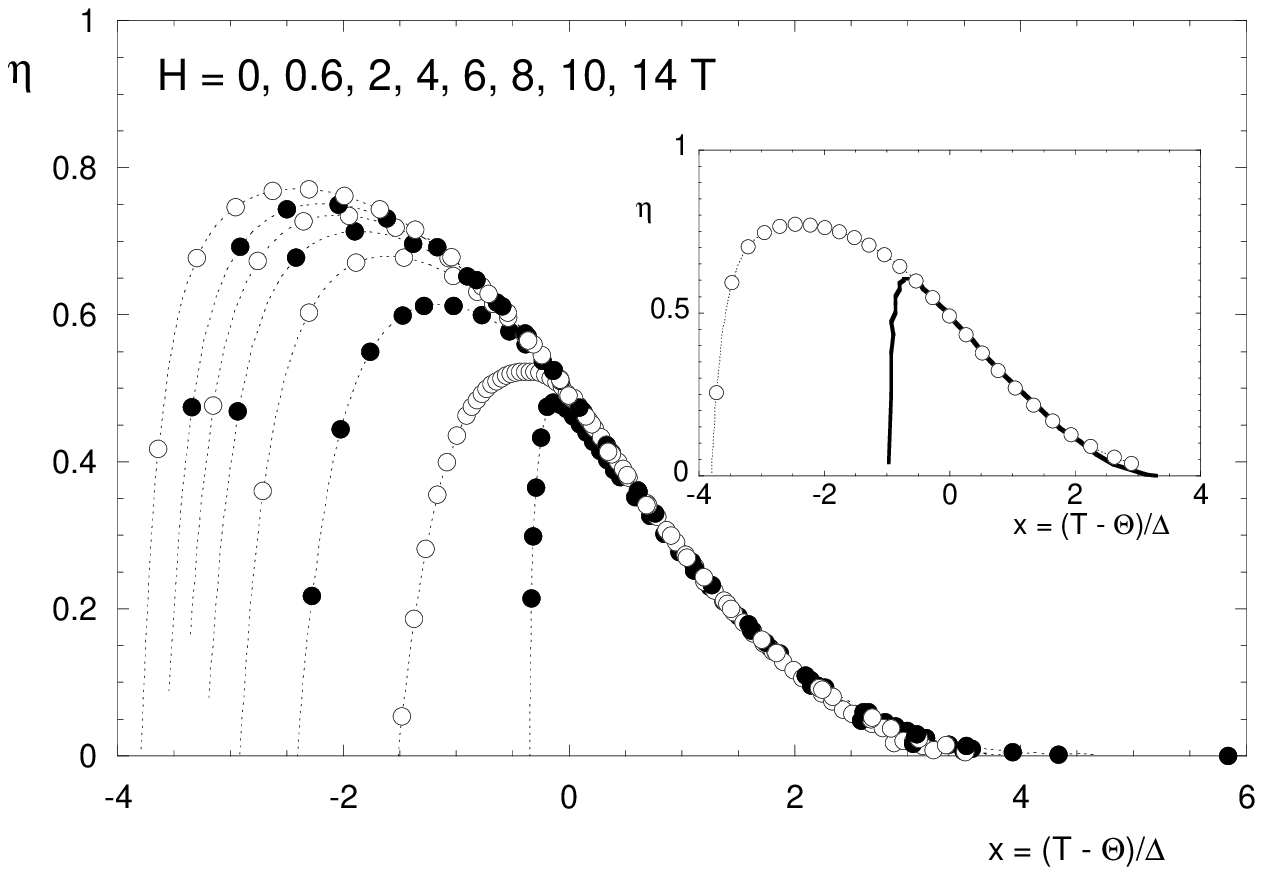}
\caption{Scaling of $\eta$ at various fields for the sample with $\delta = 0.235$. Inset: perfect collapse of the data reported in the main panel on the curve $\eta(T)$ obtained at $H=0$ for the sample with $\delta= 0.213$ (see also Fig. \ref{fig2}). Only the data at 14 T are shown, for clarity.}
\label{fig5}
\end{figure}

To support this hypothesis and to give further insight into the process of formation of preformed pairs, we now comment on the doping dependence of the two scaling parameters $\Theta$ and $\Delta^*$. In Fig.\ref{fig4} we report the values for $\Theta$ and $\Theta+\Delta^*$ in the doping-temperature phase diagram, together with $T^*$ and $T_c$. As a first result, we note that $\Theta+\Delta^*$ is nearly coincident with $T^*$. This result is obvious for the sample with $\delta = 0.22$ for which this equality is valid by definition, but is a good consistency test for all the other curves, since in the scaling procedure $\Theta$ and $\Delta^*$ are simply scaling parameters, without any {\em a priori}  relation with $T^*$. Second, we note that the linear extrapolation of the curves $T^*(\delta)$ and $\Theta(\delta)$ cross at $\delta = 0.31$. Although the initial choice for the threshold defining $\Theta$ (i.e.  $\eta(\Theta) = 0.5$) is arbitrary, by choosing different references for $\eta$ one obtains different lines $\Theta(\delta)$  which are all approximately linear and all cross the extrapolated $T^*(\delta)$ at the same value of $\delta$. This is seen in Fig.\ref{fig4}, where the dashed lines represent the curves $\Theta(\delta)$ obtained by initially choosing $\eta(\Theta) = 0.4$ and 0.3 respectively. This is a relevant result: we recall that, 
within the model for the normal state here adopted, we found that at $\delta \simeq 0.31$ the effect of the scattering by impurities becomes the same for carrier motion along and across the planes. At this doping it is possible to define a carrier velocity along the c-axis and the system becomes a 3D metal. The coincidence between the vanishing of the temperature region between $T^*$ and $\Theta$ (that is the vanishing of the crossover region) and of the difference between in-plane and out-of-plane scattering rates by impurities indicates that, in a purely 3D system, the formation of pairs and their condensation into the superconducting state occur simultaneously.\\
Finally, we come to the data in presence of an applied magnetic field. As a first result, we find that the scaling of the gapped charge carrier fraction $\eta$ is robust also with the application of the magnetic field.  We have performed measurements of $\rho_c$ in the sample with $\delta$=0.232 ($\Theta=$ 90 K, $\Delta^{*}=$ 79 K obtained in Fig.\ref{fig2}) with a magnetic field of intensity up to 14 T applied along the $c$ axis.  The gapped fraction $\eta$ coincides with the one obtained without the magnetic field (in the temperature range above $T_{c}(H=0)$).  In Fig.\ref{fig5} we report the gapped fraction as a function of the scaling variable $x$ for all the magnetic fields investigated (we stress that no additional scaling parameters are involved, since $\Theta$ and $\Delta^{*}$ are the same as in the zero-field scaling), together with the data for the most underdoped sample ($\delta$=0.213, full thick line).  The scaling on the previously obtained reference curve is again fulfilled, but the magnetic field allows for the extension of the scaling at temperatures well below the zero-field data: for $H=$ 14 T the range of the scaling is increased by approximately 22 K. Thus, the magnetic field does not influence the fraction of the gapped
carriers: the effect of the magnetic field only determines a widening of the temperature range where the ``pseudogap'' phenomenon is observed (apart an additional rounding of the transition of $\eta$ near $T_c(H)$, as in the direct measurements of $\rho_c$).  Since in the present scheme the gapped carriers tend to localize within the (a,b) planes, it comes out that the magnetic field does not affect the degree of localization: the magnetic field does not influence the number of carriers in the gapped state, while it strongly reduces the macroscopic phase coherence leading to the superconductivity.\\
In conclusion, the two barriers analysis of the c-axis resistivity for the normal state of Bi:2212 is continued below the temperature $T^*$ for a large spectrum of the doping $\delta$. We obtain the fraction $\eta(T)$ of the "gapped" carriers that do not participate to the conduction along the c-axis. We are able to show that $\eta(T)$ can be scaled on an universal curve for all $\delta$ values. The doping dependent value of $\eta$ at $T_c$ when multiplied by the activation term parameter $\beta$, proportional to the number of carriers, is constant with respect to $\delta$, indicating that the superconducting transition occurs at a doping-independent value of the "gapped" carrier density. Furthermore, the analysis of the measurements in an external magnetic field up to 14 T shows that this scaling is fulfilled at lower temperatures also as the field increases. The magnetic field does not influence the localization in the CuO planes below $T^*$ but widens the temperature range of the scaling by suppressing the superconducting phase coherence.

\newpage

\begin{references}
\bibitem{notalinear}
With respect to Ref.\cite{giura}, in Equation \ref{oldrho} we have taken as the in-plane normal state 
$\rho_{ab,n}$ the linear extrapolation obtained from the data above $T^*$, thus implicitly assigning to the pseudogap opening a possibly appearing curvature.
\end{references}
\end{document}